\documentclass[aps,prl,reprint,twocolumn,superscriptaddress,floatfix]{revtex4-2}

\usepackage{amsmath,amssymb,bm}
\usepackage{graphicx}
\usepackage{xcolor}
\usepackage[colorlinks=true,citecolor=blue,linkcolor=blue,urlcolor=blue]{hyperref}
\usepackage{microtype}
\usepackage{xspace}

\definecolor{MyGreen}{RGB}{34,139,34}

\graphicspath{{figs/}}

\newcommand{\CoSnS}{Co$_3$Sn$_2$S$_2$\xspace}
\newcommand{\EF}{E_{\mathrm F}}

\newcommand{\RZ}{\mathcal{Z}}

\begin{document}

\title{Orbital-selective correlations in kagome flat bands of Co$_3$Sn$_2$S$_2$}

\author{Rongxiang Zhu}
\affiliation{National Laboratory of Solid State Microstructures, School of Physics, Nanjing University, Nanjing 210093, China}
\affiliation{Center for Joint Quantum Studies and Department of Physics, School of Science, Tianjin University, 135 Yaguan Road, Tianjin 300350, China}

\author{Huaiqiang Wang}
\affiliation{Center for Quantum Transport and Thermal Energy Science, Institute of Physics Frontiers and Interdisciplinary Sciences, School of Physics and Technology, Nanjing Normal University, Nanjing 210023, China}
\affiliation{Jiangsu Physical Science Research Center, Nanjing 210093, China}

\author{Haijun Zhang}
\email[Contact author:]{zhanghj@nju.edu.cn}
\affiliation{National Laboratory of Solid State Microstructures, School of Physics, Nanjing University, Nanjing 210093, China}
\affiliation{Jiangsu Physical Science Research Center, Nanjing 210093, China}
\affiliation{Collaborative Innovation Center of Advanced Microstructures, Nanjing University, Nanjing 210093, China}
\affiliation{Jiangsu Key Laboratory of Quantum Information Science and Technology, Nanjing University, Suzhou 215163, China}
\affiliation{Shishan Laboratory, Suzhou Campus of Nanjing University, Suzhou 215000, China}

\author{Jianzhou Zhao}
\email[Contact author:]{jzzhao@tju.edu.cn}
\affiliation{Center for Joint Quantum Studies and Department of Physics, School of Science, Tianjin University, 135 Yaguan Road, Tianjin 300350, China}

\date{\today}

\begin{abstract}
Geometrical flat bands provide an ideal platform for studying electronic correlations and quantum-geometric effects.
Although the ferromagnetic Weyl semimetal \CoSnS is generally considered a moderately correlated system, recent spectroscopy experiments have revealed a flat-band feature near the Fermi level ($\EF$) that is absent in conventional density functional theory (DFT) calculations.
Using DFT plus dynamical mean-field theory and a minimal kagome model, we show that this feature arises not from heavy-fermion-like dynamical band renormalization, but from the shifting of a $d_{x^2-y^2}$ dominated geometrical kagome flat band toward $\EF$ driven by orbital- and spin-dependent self-energies. 
The relocated flat band enhances the local Berry curvature near $\EF$, resulting in a strongly energy-dependent anomalous Hall conductivity and a giant anomalous Nernst effect.
Our findings establish an orbital-selective correlation mechanism for positioning geometrical flat bands at $\EF$, and propose anomalous thermoelectric transport as a sensitive probe of flat band induced topology.
\end{abstract}

\maketitle

\paragraph*{\textit{Introduction.}}
Flat bands, in which the quenched kinetic energy amplifies the role of interactions, have attracted great research interest in condensed matter physics~\cite{regnaultCatalogueFlatbandStoichiometric2022}. 
Those with nontrivial topology are particularly interesting because more intriguing phenomena emerge from the interplay of quantum geometry and correlations~\cite{sunNearlyFlatbandsNontrivial2011,calugaruGeneralConstructionTopological2022,checkelskyFlatBandsStrange2024a,gaoQuantumGeometryPhenomena2026,liStableTopologyExactly2026}.
In systems with broken time-reversal symmetry, such as ferromagnets, the quantum geometry of flat-band wave functions can produce a large Berry curvature, leading to a giant anomalous Hall conductivity (AHC) and anomalous Nernst effect (ANE)~\cite{yaoFirstPrinciplesCalculation2004,xiaoBerryPhaseEffects2010,nagaosaAnomalousHallEffect2010,Li2026}.
Fundamentally, flat bands can be understood through two contrasting pictures~\cite{regnaultCatalogueFlatbandStoichiometric2022}: at one extreme, strong dynamical renormalization in heavy-fermion metals produces an extremely flat quasiparticle band and a Kondo resonance naturally tied to $\EF$~\cite{wengMultipleQuantumPhase2016,wangFlatTopologicalNodal2026,chenEmergentFlatBand2024}; at the other, destructive interference creates a geometrical flat band already at the single-particle level on frustrated lattices such as the kagome network~\cite{disanteKagomeMetals2026}.
While such geometrical flatness is robust, the band position is governed is determined by filling, crystal fields~\cite{ortizCsV$_3$Sb$_5$Z$_2$Topological2020,wilsonAV3Sb5KagomeSuperconductors2024,zhaoElectronicCorrelationsNormal2021}, and charge or magnetic ordering~\cite{jiangFeGeBuildingBlock2025,wangEnhancedSpinpolarizationPartial2023a,yiChargeDynamicsUnconventional2025}. 
In this context, ferromagnetic (FM) kagome materials are especially favorable because geometrical flat bands, broken time-reversal symmetry, spin-orbit coupling (SOC), and band topology can all coexist in the same low-energy manifold.
The shandite compound \CoSnS, in which Co atoms form a kagome lattice, develops a ferromagnetic Weyl semimetal state featuring three pairs of Weyl points and surface Fermi arcs~\cite{liuMagneticWeylSemimetal2019,moraliFermiarcDiversitySurface2019,belopolskiSignaturesWeylFermion2021,xuTopologicalSurfaceFermi2018}.
The DFT calculated large intrinsic AHC, with a characteristic scale of $10^3\ \mathrm{S/cm}$, has been confirmed experimentally~\cite{liuGiantAnomalousHall2018,Wang2018,shen33GiantAnomalous2020,guguchiaTunableAnomalousHall2020}.

Despite this established band picture, multiple spectroscopic measurements, including angle-resolved photoemission spectroscopy (ARPES)~\cite{Wang2018,Lv2025}, scanning tunneling spectroscopy (STS)~\cite{Yin2019}, and resonant inelastic X-ray scattering (RIXS)~\cite{Nag2022}, have revealed flat-band signatures near $\EF$ that are absent in conventional bulk ferromagnetic DFT calculations.
The origin of these signatures remains under debate: while STS measurements~\cite{Yin2019,xingLocalizedSpinorbitPolaron2020,huangDiscoveryConstructionSurface2023} have been interpreted as surface states, RIXS~\cite{Nag2022} and the pronounced temperature dependence of the AHC~\cite{Lv2025} suggest a bulk origin.
Crucially, experimental renormalization factors of $0.7\sim0.8$ deduced from both optical conductivity~\cite{Xu2020,yangMagnetizationInducedBandShift2020} and ARPES~\cite{liuMagneticWeylSemimetal2019} indicate only moderate electronic correlations in \CoSnS, ruling out the strong dynamical mass enhancement characteristic of heavy-fermion systems.
This raises a fundamental question: how can moderate correlations bring a geometrical kagome flat band close to $\EF$ without heavy-fermion-like mass renormalization?

In this work, we combine DFT plus dynamical mean-field theory (DMFT) with a minimal kagome model to uncover the mechanism underlying the near-$\EF$ flat band in \CoSnS. 
We demonstrate that orbital- and spin-dependent static self-energies shift a $d_{x^2-y^2}$-dominated flat band toward $\EF$, while the calculated quasiparticle weights confirm only moderate band renormalization.
A minimal kagome model captures how this orbital-selective shift brings the frustrated flat band to the Fermi level. 
Furthermore, the relocated flat band substantially enhances the local Berry curvature via extended anticrossings. 
While the integrated AHC remains in good agreement with experiment, its steep energy dependence significantly enhances the low-temperature anomalous thermoelectric response. 
This quantitatively reproduces experimental observations, providing a sensitive test of the proposed mechanism.
Computational details are provided in the Supplemental Material~\cite{sm}.

\begin{figure}[tb]
 \centering
 \includegraphics[width=0.98\columnwidth]{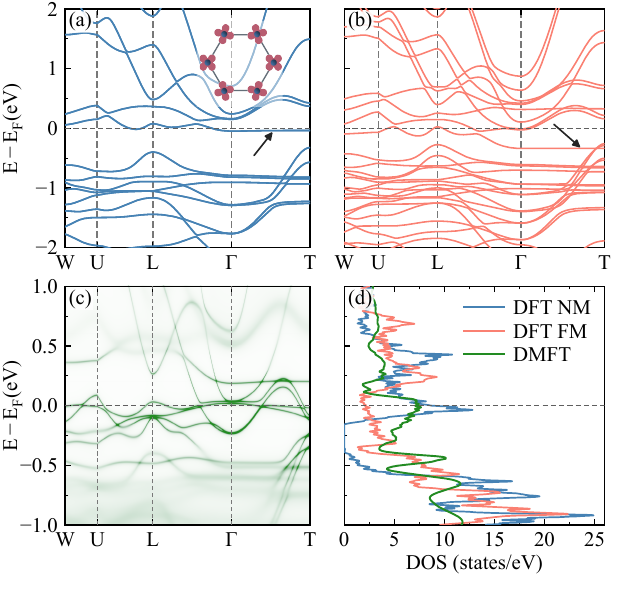}
\caption{Low-energy electronic structure. (a) NM DFT band structure with SOC. The charge density of the flat band at $\EF$ is illustrated on the top right. (b) FM DFT band structure with SOC. (c) Momentum-resolved DFT+DMFT spectral function in the FM state. (d) Total density of states obtained from NM DFT, FM DFT, and DFT+DMFT.}
\label{fig:electronic}
\end{figure}

\paragraph*{\textit{Coherent flat band.}}

We first examine how ferromagnetism and electronic correlations reshape the low-energy kagome bands in \CoSnS. Figure~\ref{fig:electronic} compares the electronic structures obtained from nonmagnetic (NM) DFT, FM DFT, and DFT+DMFT. In the NM band structure [Fig.~\ref{fig:electronic}(a)], a weakly dispersive band with $d_{xy}$ character already appears near $\EF$. 
Kagome destructive interference supplies this low-dispersion manifold in the parent NM structure, whose large density of states (DOS) strongly favors an itinerant Stoner ferromagnetic instability. Once FM exchange is introduced at the DFT level [Fig.~\ref{fig:electronic}(b)], the spin degeneracy is lifted. The resulting spin splitting substantially reconstructs the low-energy electronic structure, suppressing the flat-band contribution at the Fermi level and pushing the localized kagome states well below $\EF$.

The momentum-resolved DFT+DMFT spectral function in Fig.~\ref{fig:electronic}(c) qualitatively changes this picture: a sharp, weakly dispersive feature appears near $\EF$. Despite its narrow bandwidth, the feature remains remarkably coherent, as the imaginary part of the low-frequency Matsubara self-energy extrapolates to zero as $\omega_n \to 0$~\cite{sm}, confirming Fermi-liquid-like behavior with well-defined quasiparticles.
Our DFT+DMFT results capture the key features of the recent ARPES measurements~\cite{Lv2025, sm}, particularly the nearly dispersionless band near $\EF$. As shown below, this emergent feature possesses predominantly $d_{x^2-y^2}$ character, distinguishing it from the $d_{xy}$-dominated flat band in the NM calculation.
Correspondingly, the total DOS in Fig.~\ref{fig:electronic}(d) reveals that the value at $\EF$ is enhanced by a factor of two to three compared with the FM DFT result, providing a direct signature of the correlation-induced enhancement of low-energy spectral weight near the Fermi level.

\begin{figure}[tb]
 \centering
 \includegraphics[width=0.98\columnwidth]{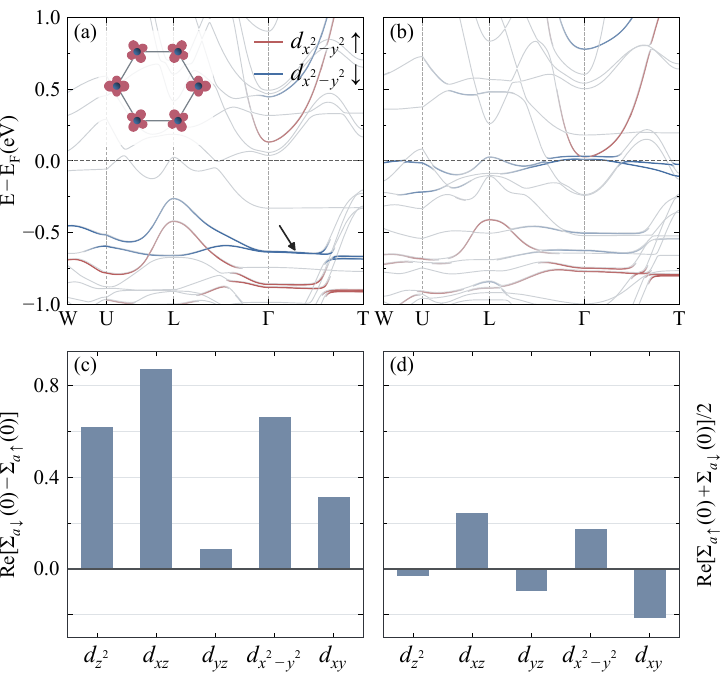}
\caption{Quasiparticle band structure and self-energy decomposition. (a) FM DFT+SOC bands projected onto the Co $d_{x^2-y^2}$ orbital. Red and blue denote the spin-up and spin-down $d_{x^2-y^2}$ contributions, respectively. (b) The corresponding quasiparticle bands from Eq.~(\ref{eq:hqp}). (c) Orbital dependent spin-antisymmetric zero-frequency self-energy $\Delta\Sigma_a=\Re[\bar{\Sigma}_{a\downarrow}(0)-\bar{\Sigma}_{a\uparrow}(0)]$. (d) Spin-averaged component $\Sigma_{c,a}=\tfrac{1}{2}\Re[\bar{\Sigma}_{a\uparrow}(0)+\bar{\Sigma}_{a\downarrow}(0)]$.}
\label{fig:selfenergy}
\end{figure}

\paragraph*{\textit{Orbital-selective correlation.}}
To understand the origin of the low-energy feature, we construct the quasiparticle (QP) Hamiltonian from the DFT+DMFT self-energy
\begin{equation}
 H_{\mathrm{qp}}=\RZ^{1/2}\left[H_0+\Re\bar{\Sigma}(0) - \mu \right]\RZ^{1/2}
 \label{eq:hqp}
\end{equation}
Here, $\RZ = \left[1-\left.\frac{\partial \Im\Sigma(\mathrm{i}\omega)}{\partial \omega}\right|_{\omega\rightarrow 0^+}\right]^{-1}$ is the QP weight which characterizes correlation-induced band renormalization, $H_0$ is the spin-polarized Wannier Hamiltonian obtained from DFT, and $\bar{\Sigma}(0) \equiv \Sigma(0) - \Sigma_{\mathrm{dc}}$ is the self-energy after subtraction of the double-counting correction. The chemical potential is denoted by $\mu$. The QP band structure presented in Fig.~\ref{fig:selfenergy}(b) reproduces the low-energy feature obtained from DFT+DMFT~\cite{sm}.
Compared to the DFT Hamiltonian $H_0$, $H_{\mathrm{qp}}$ contains two correlation-induced effects: the renormalization factor $\RZ$ and the zero-frequency self-energy $\Re\bar{\Sigma}(0)$.
From the continuous-time quantum Monte Carlo (CTQMC) self-energy, the renormalization factors $\RZ$ range from 0.74 to 0.82, corresponding to moderate correlations and agreeing with estimates from optical conductivity and ARPES~\cite{Xu2020,yangMagnetizationInducedBandShift2020}. Although this moderate band narrowing affects the final QP dispersion~\cite{sm}, it cannot account for the several-hundred-meV shift between Figs.~\ref{fig:selfenergy}(a) and (b).
Instead, the pronounced energy shift of the flat-band branch is driven predominantly by the static self-energy $\Re\bar{\Sigma}(0)$.
To trace the band structure evolution, Figs.~\ref{fig:selfenergy}(a) and \ref{fig:selfenergy}(b) show the Co-$d_{x^2-y^2}$ projections of the DFT and QP band structures, respectively (projections onto the other Co-$3d$ orbitals are provided in the Supplemental Material~\cite{sm}).
In the QP band structure shown in Fig.~\ref{fig:selfenergy}(b), the flat band near $\EF$ derives primarily from the spin-down $d_{x^2-y^2}$ branch, which lies roughly 0.6 eV below $\EF$ in the FM DFT band structure in Fig.~\ref{fig:selfenergy}(a).
Electronic correlations thus substantially amplify the splitting between the spin-up and spin-down $d_{x^2-y^2}$ manifolds.

To further shed light on the microscopic mechanism underlying this shift, we decompose the zero-frequency self-energy into a spin-antisymmetric component, $\Delta\Sigma_a(0)=\Re[\bar{\Sigma}_{a\downarrow}(0)-\bar{\Sigma}_{a\uparrow}(0)]$, and a spin-averaged component, $\Sigma_{c,a}(0)=\tfrac{1}{2}\Re[\bar{\Sigma}_{a\uparrow}(0)+\bar{\Sigma}_{a\downarrow}(0)]$.
As shown in Fig.~\ref{fig:selfenergy}(c), the spin-antisymmetric component $\Delta\Sigma_a(0)$ is strongly orbital-dependent, ranging from approximately $0.1$ eV to $0.9$ eV, with $\Delta\Sigma_{x^2-y^2}\simeq0.66$ eV.
This pronounced orbital differentiation is attributed to orbital- and spin-dependent differences in hybridization and occupation: for the $d_{x^2-y^2}$ orbital, weaker low-frequency hybridization in the spin-down channel yields a reduced spin-down occupation and an enhanced orbital spin polarization in the self-consistent DMFT calculation. In contrast, the $d_{yz}$ orbital remains nearly fully occupied with minimal spin imbalance, leading to a much smaller self-energy splitting.
The spin-averaged component $\Sigma_{c,a}(0)$ shown in Fig.~\ref{fig:selfenergy}(d) produces an additional relative orbital shift: the $d_{x^2-y^2}$ energy level is raised by approximately $0.17$ eV, whereas the $d_{xy}$ orbital level is pushed downward. Together, the enhanced effective exchange splitting and the upward shift of the orbital center bring the spin-down $d_{x^2-y^2}$ geometrical flat band toward $\EF$.

\begin{figure}[tb]
 \centering
 \includegraphics[width=0.99\columnwidth]{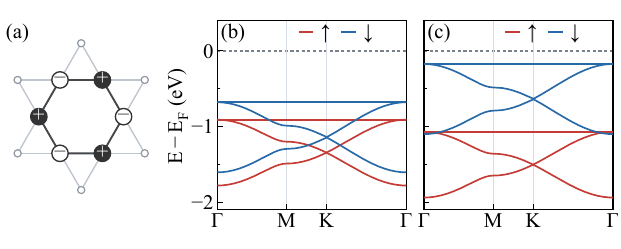}
\caption{Flat-band formation and relocation. (a) Schematic illustration of kagome flat-band formation. The filled and open circles with $+$ and $-$ signs represent the alternating wave-function phases. (b) Band structure of the three-band Slater-Koster model $H_{\mathrm{SK}}$ with nearest-neighbor hopping. Red and blue solid curves denote the spin-up and spin-down channels, respectively.  (c) Band structure obtained by adding the static DMFT self-energy correction.}
\label{fig:model}
\end{figure}

\paragraph*{\textit{Minimal kagome model.}}

To illustrate the flat-band relocation mechanism, we construct a nearest-neighbor kagome model from the FM Wannier Hamiltonian. 
Retaining the DFT spin-dependent onsite energies and neglecting explicit SOC, the minimal model reads
\begin{equation}
 H_{\mathrm{SK}}
 =\sum_{is}\varepsilon_{ds}^{\mathrm{DFT}}n_{is}
 +\sum_{\langle ij\rangle s}t_{\mathrm{SK},s}
 \left(c^\dagger_{is}c_{js}+\mathrm{H.c.}\right).
\label{eq:kagome}
\end{equation}
Here, $s=\uparrow,\downarrow$ labels spin, $c^\dagger_{is}$ creates a spin-$s$ electron on kagome site $i$, and $n_{is}=c^\dagger_{is}c_{is}$. The onsite energies $\varepsilon_{ds}^{\mathrm{DFT}}$ retain the DFT spin splitting of the selected orbitals. For the present geometry, $t_{\mathrm{SK},s}=3V_{dd\sigma}^{s}/16-3V_{dd\pi}^{s}/4+V_{dd\delta}^{s}/16$, where $V_{dd\sigma}^{s}$, $V_{dd\pi}^{s}$, and $V_{dd\delta}^{s}$ are the effective spin-dependent two-center hopping integrals.
Details of the SK fitting procedure and comparison with the full Wannier Hamiltonian are provided in the Supplemental Material~\cite{sm}.
The geometrical origin of the kagome flat band is illustrated in Fig.~\ref{fig:model}(a). Equal-magnitude wave-function amplitudes with alternating signs around a kagome hexagon form a compact localized state: hopping contributions to any adjacent site outside the loop interfere destructively, yielding $t_{\mathrm{SK},s}(+1-1)=0$~\cite{wuFlatBandsWigner2007}. Consequently, the nearest-neighbor model hosts a flat band in each spin channel [Fig.~\ref{fig:model}(b)].

In the minimal model, the static self-energy correction modifies the onsite energies according to
\begin{equation}
 \varepsilon_{ds}^{\mathrm{eff}}
 =\varepsilon_{ds}^{\mathrm{DFT}}
 +\Re\bar{\Sigma}_{x^2-y^2,s}(0).
\label{eq:modelcorr}
\end{equation}
The corresponding flat-band energy becomes $E_{\mathrm{flat},s}=\varepsilon_{ds}^{\mathrm{eff}}-2t_{\mathrm{SK},s}$. 
As shown in Fig.~\ref{fig:model}(c), the static correction shifts the spin-down flat band from approximately $-0.679$ eV to $-0.173$ eV, bringing it closer to $\EF$ while preserving its geometrical flatness.
It should be noted that in the realistic material, longer-range hopping, interlayer coupling, SOC-induced spin mixing, and hybridization with other orbitals further modify the flat-band energy and induce a weak dispersion. Thus, the minimal model captures the essential relocation mechanism, whereas the full DFT+DMFT calculation places the weakly dispersive branch in close proximity to $\EF$.
This mechanism differs fundamentally from the heavy-fermion scenario, in which strong dynamical renormalization produces a flat quasiparticle band that is pinned to $\EF$ by a Kondo resonance.

\begin{figure}[tb]
 \centering
 \includegraphics[width=0.98\columnwidth]{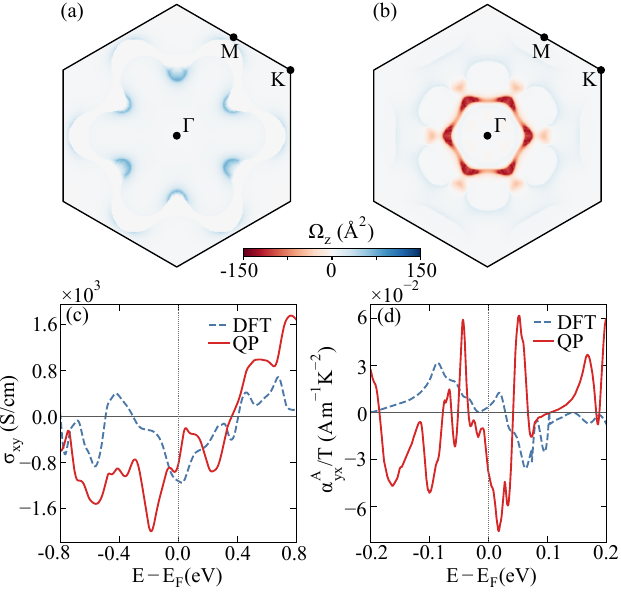}
\caption{Correlation-induced Berry-curvature redistribution and anomalous transverse response. (a) Berry curvature in the $k_z=0$ plane for FM DFT summed over all occupied
 states. (b) Berry curvature in the $k_z=0$ plane for the DMFT quasiparticle Hamiltonian. (c) Intrinsic anomalous Hall conductivity $\sigma_{xy}$ as a function of the chemical potential $\mu$. (d) Low-temperature anomalous thermoelectric conductivity $\alpha_{yx}^{A}/T = - \alpha_{xy}^{A}/T$ obtained from the Mott relation.}
\label{fig:berry}
\end{figure}

\paragraph*{\textit{Berry curvature and anomalous thermoelectric response.}}
We next examine how flat-band relocation affects Berry curvature and anomalous transverse transport.
The band-resolved Berry curvature is given by
\begin{equation}
    \Omega^z_n(\mathbf{k})=-2\hbar^2\sum_{m\neq n}\frac{\Im\langle \psi_{n\mathbf{k}}|\hat{v}_x|\psi_{m\mathbf{k}}\rangle\langle \psi_{m\mathbf{k}}|\hat{v}_y|\psi_{n\mathbf{k}}\rangle}{(\varepsilon_{m\mathbf{k}}-\varepsilon_{n\mathbf{k}})^2},
 \label{eq:berry}
\end{equation}
where $|\psi_{n\mathbf{k}}\rangle$ is the Bloch eigenstate of band $n$ at momentum $\mathbf{k}$, $\hat{v}_x$ and $\hat{v}_y$ are the corresponding velocity operators~\cite{thoulessQuantizedHallConductance1982}.
Owing to the energy denominator in Eq.~(\ref{eq:berry}), small interband gaps naturally produce concentrated Berry curvature.
In \CoSnS, the correlation-induced flat band near $\EF$ produces closely spaced bands and associated anticrossings, which can therefore enhance the Berry curvature.
In the $k_z=0$ plane, the occupied Berry curvature in the FM DFT calculation exhibits localized peaks with a maximum absolute value of 77 \AA$^2$ in Fig.~\ref{fig:berry}(a). In contrast, the QP result generates a pronounced sixfold-symmetric ring around $\Gamma$ with the peak value reaching 129 \AA$^2$ as shown in Fig.~\ref{fig:berry}(b).
The weak dispersion of the correlation-induced flat band enhances the Berry curvature over an extended region of the Brillouin zone, rather than only near isolated points as in the DFT result.
Notably, the relocated flat band coexists with topological Weyl nodes approximately 38 meV above $\EF$, whose topological character is confirmed by Wilson-loop calculations and further supported by the corresponding surface Fermi arcs and Berry-curvature monopoles~\cite{sm}.

The intrinsic AHC can be obtained by integrating the Berry curvature over the Brillouin zone:
\begin{equation}
 \sigma_{xy}(\mu,T)=-\frac{e^2}{\hbar}\sum_n\int_{\mathrm{BZ}}
 \frac{d^3k}{(2\pi)^3}f(\varepsilon_{n\mathbf{k}}-\mu,T)
 \Omega^z_n(\mathbf{k}).
 \label{eq:ahc}
\end{equation}
where $f$ is the Fermi-Dirac distribution, while $\varepsilon_{n\mathbf{k}}$ and $\Omega^z_n(\mathbf{k})$ are the energy and Berry curvature of band $n$, respectively. We evaluate the Brillouin-zone integral using a dense $200\times200\times200$ $k$ mesh.
Despite the enhancement of the local Berry curvature, the DFT and QP calculations yield comparable AHCs at $\EF$, both of which are consistent with the experimental value of approximately $900\ \mathrm{S/cm}$~\cite{liuMagneticWeylSemimetal2019,Wang2018,Lv2025}.
An enhanced local Berry curvature does not necessarily imply a larger integrated AHC, because contributions with opposite signs can cancel upon integration over the Brillouin zone.

Although the DFT and QP calculations yield similar AHC values at $\EF$, the QP result exhibits a substantially stronger energy dependence of $\sigma_{xy}(\mu)$. As shown in Fig.~\ref{fig:berry}(c), the QP curve develops a pronounced peak at $\mu\simeq-0.05$ eV, which is absent in the DFT calculation, resulting in a substantially enhanced energy derivative at $\EF$. 
At low temperatures, this energy derivative directly dictates the anomalous thermoelectric conductivity (anomalous Nernst response) via the Mott relation:

\begin{equation}
 \frac{\alpha^A_{xy}}{T}=-\frac{\pi^2k_\mathrm{B}^2}{3e}
 \left.\frac{\partial\sigma_{xy}}{\partial\varepsilon}\right|_{E_\mathrm{F}}.
 \label{eq:mott}
\end{equation}
Experimentally, $|\alpha^A_{xy}|/T$ ranges from $0.05$ to $0.10\ \mathrm{A\,m^{-1}\,K^{-2}}$ across different samples~\cite{Ding2019}.
Our QP calculation yields $|\alpha^A_{xy}|/T\simeq0.05\ \mathrm{A\,m^{-1}\,K^{-2}}$, which lies within the experimental range and agrees more closely with experiment than the DFT result [Fig.~\ref{fig:berry}(d)]. This improved agreement supports the role of correlation-induced flat band relocation in the anomalous thermoelectric response.

\paragraph*{\textit{Discussion.}}
Despite extensive attention devoted to \CoSnS as a prototypical ferromagnetic Weyl semimetal, the microscopic origin and bulk or surface character of its near-$\EF$ flat-band signatures remain under debate. Here, we establish a bulk mechanism based on the correlation-induced orbital-selective relocation of a geometrical kagome flat band.
Unlike the heavy-fermion scenario, neither the flat band nor its associated Berry curvature enhancement and anomalous thermoelectric response arise from extreme dynamical mass renormalization. Instead, they result from the combination of lattice geometry, orbital-selective correlations, and spin-orbit coupling.
Specifically, this mechanism proceeds through three steps: first, kagome destructive interference generates the $d_{x^2-y^2}$ dominated flat band; second, the orbital and spin dependent zero-frequency self-energies shift the relevant branch toward $\EF$; and third, SOC generates extended anticrossings near $\EF$, which are strong sources of Berry curvature. These structural, many-body, and topological components are together account for the observed phenomenology.
The agreement between the calculated spectral function and the near-$\EF$ ARPES feature~\cite{Lv2025} supports this bulk flat-band relocation mechanism.

Although the flat-band relocation is governed primarily by the zero-frequency real part of the self-energy, this does not imply that a static mean-field treatment is sufficient to reproduce the same electronic structure. Within DMFT, this shift is obtained from a self-consistent solution of the interacting impurity problem, which includes local quantum fluctuations and coupling to a dynamical bath. In contrast, a static mean-field treatment such as DFT+\textit{U} does not capture dynamical screening and fluctuations.
As a comparison, our DFT+\textit{U} calculations using the identical interaction parameters fail to reproduce either the low-energy flat-band feature obtained in DFT+DMFT or the feature observed in ARPES~\cite{sm}. This clear contrast demonstrates that a dynamical many-body treatment of correlations is crucial for capturing the low-energy electronic structure of \CoSnS.

Very recently, an STS study reported a zero-bias peak on the reconstructed Co$_3$Sn surface of \CoSnS, attributed to a higher-order van Hove singularity of a surface band~\cite{kumarnagPomeranchukInstabilityInduced2026}. 
While our proposed bulk mechanism does not exclude additional surface contributions, these findings highlight the need to distinguish bulk and surface contributions experimentally.
High-resolution photon-energy-dependent ARPES could help resolve this issue by probing the out-of-plane dispersion of the near-$\EF$ feature and comparing the measured spectra with the calculated bulk bands.
Complementary evidence could be obtained from the anomalous thermoelectric response. Our calculated $\sigma_{xy}(\mu)$ suggests that modest chemical-potential shifts can substantially change $\alpha^A_{xy}/T$, while producing a more moderate variation of the AHC near $\EF$. Controlled carrier doping in bulk crystals, combined with ARPES and thermoelectric measurements, could therefore clarify whether the flat-band position and thermoelectric response evolve together as predicted.
Furthermore, the enhanced DOS from this near-$\EF$ flat band could substantially couple to chiral phonons reported recently in \CoSnS~\cite{cheMagneticOrderInduced2025,yangInherentCircularDichroism2025,wangInitioTheoryPhonon2025}, opening possibilities for exploring chiral electron-phonon coupling.
Together, our work establishes an orbital-selective correlation mechanism that shifts the geometrical flat bands toward the Fermi level, providing new insight into the interplay between electronic correlations and band topology in quantum materials.

\begin{acknowledgments}
We thank Jiaxin Yin for very helpful discussions. 
This work was supported by the National Natural Science Foundation of China (Grants No. 12474239, No. 12534007, and No. 92365203), the National Key Research and Development Program of China (Grants No. 2024YFA1409100), the National Key Laboratory of Shock Wave and Detonation Physics Fund (Grants No. JCKYS2025212108), the Natural Science Foundation of Jiangsu Province (Grants No. BK20253012, No. BK20252117, No. BK20243011, and No. BK20233001), the Fundamental Research Funds for the Central Universities (Grants No. KG202501), and the Fundamental and Interdisciplinary Disciplines Breakthrough Plan of the Ministry of Education of China (Grant No. JYB2025XDXM409).
The computational work was performed on the resources of the e-Science Center of Collaborative Innovation Center of Advanced Microstructures.
\end{acknowledgments}

\bibliography{refs}

\end{document}